\documentclass[iop]{emulateapj}

\usepackage{amsmath}
\def\bwt{\begin{widetext}}
\def\ewt{\end{widetext}}
\def\beq{\begin{equation}}
\def\eeq{\end{equation}}

\def\bY{{\rm Y}}
\def\bPsi{{\rm \Psi}}
\def\bPhi{{\rm \Phi}}

\newcommand{\bea}{\begin{eqnarray}}
\newcommand{\eea}{\end{eqnarray}}
\def\bmat{\begin{displaymath}}
\def\emat{\end{displaymath}}
\def\bear{\begin{eqnarray}}
\def\eear{\end{eqnarray}}
\usepackage{color}

 \usepackage[colorlinks]{hyperref}

\usepackage[normalem]{ulem}

\slugcomment{Submitted to Astrophysical Journal}

\shorttitle{The Mass and Radii of Strongly Magnetized Neutron Stars}
\shortauthors{Kamiab, Broderick, and Afshordi}

\begin{document}


\title{The Mass and Radii of Strongly Magnetized Neutron Stars}


\author{Farbod Kamiab, Avery E. Broderick, and Niayesh Afshordi}
\affil{Perimeter Institute for Theoretical Physics, 31 Caroline Street North, Waterloo, ON N2L 2Y5, Canada\\
Department of Physics and Astronomy, University of Waterloo, 200 University Avenue West, Waterloo, ON, N2L 3G1, Canada}
\email{fkamiab@pitp.ca}

%


\begin{abstract}
It has been clear for some time now that super-critical surface magnetic fields, exceeding $4\times10^{13}$ G, exist on a subset of neutron stars.  These magnetars may harbor interior fields many orders of magnitude larger, potentially reaching equipartition values.  However, the impact of these strong fields on stellar structure has been largely ignored, potentially complicating attempts to infer the high density nuclear equation of state.  Here we assess the effect of these strong magnetic fields on the mass-radius relationship of neutron stars.  We employ an effective field theory model for the nuclear equation of state that includes the impact of hyperons, anomalous magnetic moments, and the physics of the crust.  We consider two magnetic field geometries, bounding the likely magnitude of the impact of magnetic fields: a statistically isotropic, tangled field and a force-free configuration.  In both cases even equipartition fields have at most a 30\% impact on the maximum mass.  However, the direction of the effect of the magnetic field depends on the geometry employed -- force-free fields leading to reductions in the maximum neutron star mass and radius while tangled fields increase both -- challenging the common intuition in the literature on the impact of magnetic fields.
\end{abstract}



\keywords{Neutron Stars, Pulsars, Magnetic Field, Magnetar}


\section{Introduction}

Observations and theoretical studies of soft gamma-ray repeaters and  X-ray pulsars point to the existence of neutron stars with very high surface magnetic fields ($B>10^{14}$ G), comprising the so-called magnetars
\citep{duncan92, Paczynski1992, thompson95, thompson96, melatos99}. 
These surface magnetic fields are inferred through the observed slowing of the stellar rotation, presumed to be a result of the emission of energy and angular momentum via large-scale magnetic fields at the light cylinder, the point beyond which they are unable to continue to rigidly rotate with the star.  This is expected to spin the star down on a timescale $\approx P/\dot{P} \propto B^{-2} P^4$, where $P$ is the spin period.  Thus, magnetars are universally observed to have long periods, roughly 1~s, and thus correspondingly large light cylinders $c P/(2\pi)\approx 5\times10^4$~km.  As a result, the implied surface fields necessarily rely on a significant extrapolation, and typically assume a dipolar magnetospheric magnetic field geometry, necessarily producing a lower limit on the surface field strength, which is itself likely to be a lower limit on the interior field strengths.

 As a recent example, \citet{spindown} monitored the temporal and spectral evolution of a pulsar, originally discovered by the NuSTAR X-ray Observatory, and from the spin-down measurement, inferred a {\it dipole} magnetic field strength $B = 3 \times 10^{14}$ G.  Magnetars can also be observed in the radio band. Follow-up observations of the pulsar PSR J1622-4950, discovered by \citet{radioloud} in a survey of radio pulsars with the Parkes 64 m telescope, show that the pulsar has the highest inferred surface magnetic field of the known radio pulsars ($B \sim 3 \times 10^{14}$ G), making it the first magnetar discovered via its radio emission. A catalog of 26 currently known magnetars was presented recently by \citet{kaspicatalogue}.

The existence of extremely strong magnetic fields observed in magnetars can be explained by a number of processes. Neutron stars with strong magnetic dipole fields $B \sim 10^{14}$ - $10^{15}$ G, can form when conditions for efficient helical dynamo action are met during the first few seconds after gravitational collapse \citep{duncan92}. In addition to differential rotation, convection may play a significant role in amplifying the magnetic field \citep{thompson1993}. In the context of more sophisticated simulations of field amplification in non-rotating stellar cores during the collapse and post-bounce accretion phases of a supernova, including magnetohydrodynamics and neutrino transport, it was found that initial magnetic field strengths stronger than $10^{10}$ G could yield magnetar-like final field strengths \citep{equipartition1}.

As neutron stars with very strong magnetic fields appear to exist in nature, one is tempted to ask how these field strengths affect the structure of these stars. This is particularly important as the equilibrium mass and radius of neutron stars vary based on the nuclear and gravitational physics assumed (for example, see \cite{lattimer2001, lattimer2007} for the effects of various nuclear equations of state on the structure of neutron stars, and see \cite{kamiab2011, pani2011, alavirad2013} for effects of modifying general relativity). As a result, observational measurements of these masses and radii have the potential to constrain theoretical models. In particular, observing neutron stars with very high masses is useful, as each set of models (nuclear equation of state and gravitational model) predicts a maximum mass beyond which no neutron stars would exist. For example, the detection of a $1.97\pm 0.04$ M$_\odot$ pulsar by \citet{demorest2010}, or the measurement of a $2.01\pm 0.04$ M$_\odot$ pulsar by \citet{antoniadis2013} have been used to significantly constrain the viable nuclear equations of state, as well as potential modifications of general relativity \citep{kamiab2011}. Therefore, before reaching definite conclusions, it is crucial to include all the necessary physics, and in the case of magnetars, investigate how strong magnetic fields affect their mass-radius (M-R) distribution.

The presence of a strong magnetic field in a neutron star can potentially affect the mass and radius in two ways:
\begin{enumerate}
 \item Locally, a magnetic field affects the nuclear equation of state (EOS) thus indirectly affecting the equilibrium configurations of the star. 
 \item The magnetic field can affect the structure of the neutron star globally by contributing to the hydrostatic support against gravity via its stress, and the structure of its spacetime via its energy (in other words, through the contribution of the Maxwell energy-momentum tensor to the Einstein equations). 
\end{enumerate}

\citet{magneticEOS, magneticEOS2} studied the effects of a strong magnetic field on the nuclear EOS. As this work is an extension of their studies, we will describe this magnetized nuclear EOS in more detail in Section \ref{secEOS}.  The EOS expresses nuclear pressure in terms of the local magnetic field strength and nuclear density. Therefore, some assumption about the structure of the magnetic field is necessary. \citet{cardall2001} studied static neutron stars with poloidal magnetic fields and a simple class of electric current distributions consistent with the requirement of stationarity. Considering the global effect of the magnetic field stress in the Einstein equations, and assuming a set of nuclear EOS, mainly the one calculated by  \citet{magneticEOS, magneticEOS2}, they found that the magnetic field increases noticeably the maximum mass. In a recent study, \citet{astashenok2014} have considered neutron stars with strong magnetic fields (where the field strength is a simple parametric function of baryon density only) in the framework of $f(R)$ gravity. Using a nuclear EOS similar to the one calculated by \citet{magneticEOS, magneticEOS2}, and assuming the global effect of the magnetic field pressure, they found that the strong magnetic field can increase considerably the maximal mass of the star. They also find that for large fields, the M-R relation differs considerably from that of general relativity only for stars with masses close to the maximal one. In another recent study, \citet{lopes2014} assume a chaotic magnetic field model introducing a variable magnetic field, which depends on the energy density rather than on the baryonic density, and based on this calculate the mass radius relationship for neutron stars.

It is important to stress that in the studies of \citet{lopes2014}, \citet{astashenok2014} and \citet{cardall2001}, the magnetic field is able to provide hydrostatic support in the global sense mentioned above where the magnetic field contributes directly in pressure in the Einstein equations [for example, $P_B={B^2/ 8\pi}$ in the case of \citet{astashenok2014} and $P_B={B^2/ 24\pi}$ in the case of \citet{lopes2014}].  Although the local effects of the magnetic field on the EOS should be considered in calculating the M-R relations, it is not certain whether the magnetic field can provide global hydrostatic support for the star. Using the principle of conservation of total helicity, \citet{narayan2008} developed a variational principle for computing the structure of the magnetic field inside a conducting sphere surrounded by an insulating vacuum. They show that, for a fixed total helicity, the minimum energy state corresponds to a {\it force-free} configuration, which is generically anisotropic. If magnetic field lines rearrange to a force-free configuration in neutron stars, then by definition,  they cannot hydrostatically support the star, and thus their impact is limited to local effects on EOS. 

The purpose of this work is to study the effect of these various assumptions \cite[in particular, the force-free model of][]{narayan2008} on the M-R distribution of neutron stars. For the effects of the magnetic field on the nuclear EOS, we base our work on the calculations of \citet{magneticEOS, magneticEOS2} which will be briefly described in Section \ref{secEOS}. In Section \ref{secB} we will motivate the particular magnetic field profile we consider, based on a simplified model of neutron star formation and magnetic field amplification that results in a local magnetic field strength fixed to a fraction of the equipartition value.  By varying the fraction of equipartition we are able to study the effect of different field strengths on the M-R distribution. In Section \ref{secMR} we will calculate the M-R relation of neutron stars, assuming different strengths of a statistically isotropic magnetic field, based on different models for the nuclear EOS. In Section \ref{forcefree} we will study the structure of neutron stars with a force-free magnetic field profile (which is anisotropic). Finally, we will conclude in Section \ref{conclude}.

\section{Magnetized Nuclear Equation of State}
\label{secEOS}
\begin{figure}
\includegraphics[angle=0,scale=0.7]{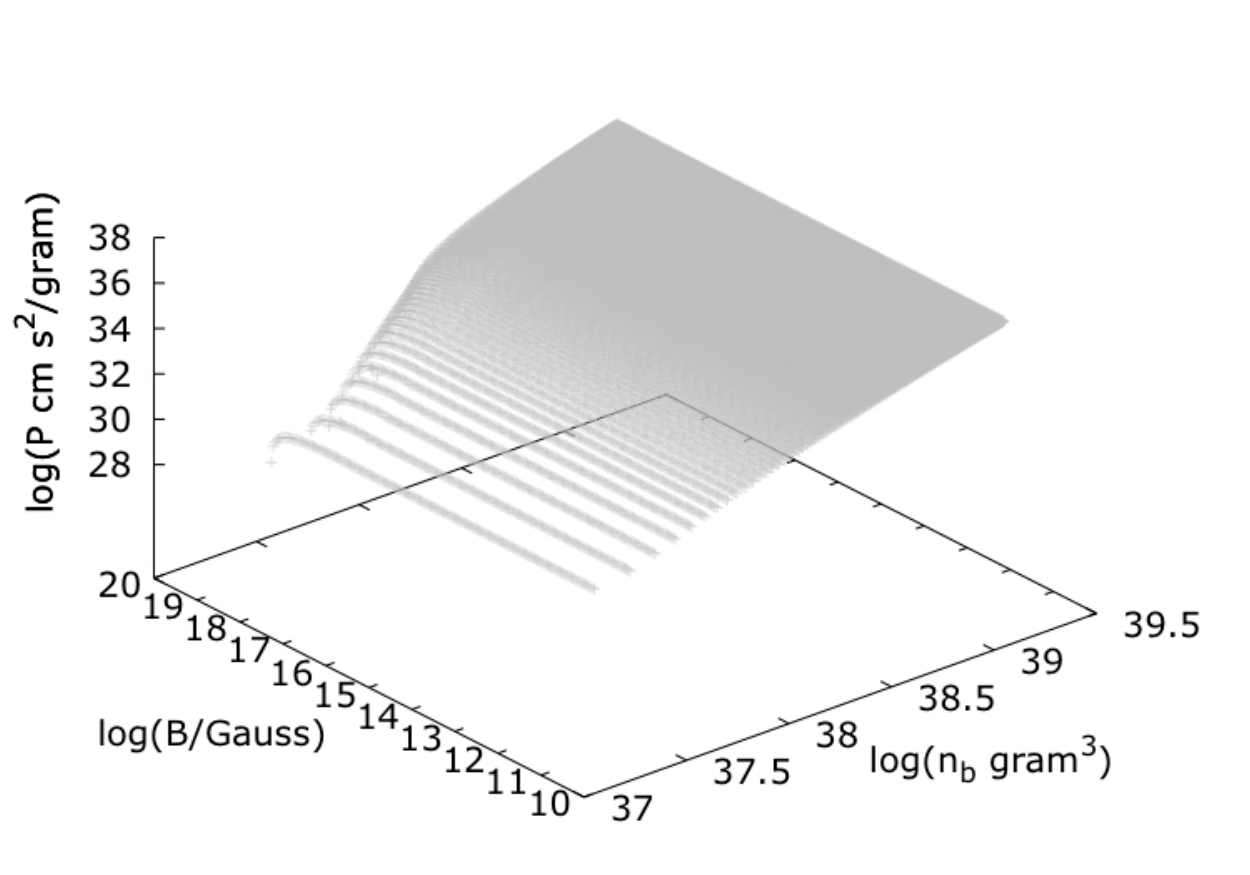}
\caption{EOS - pressure ($P$) versus number density of baryons ($n_b$) for different values of the magnetic field strength ($B$) assuming no anomalous magnetic moments. Strong magnetic fields soften the EOS because of Landau quantization.}
\label{EOS30000cgs}
\end{figure}

\begin{figure}
\includegraphics[angle=0,scale=0.7]{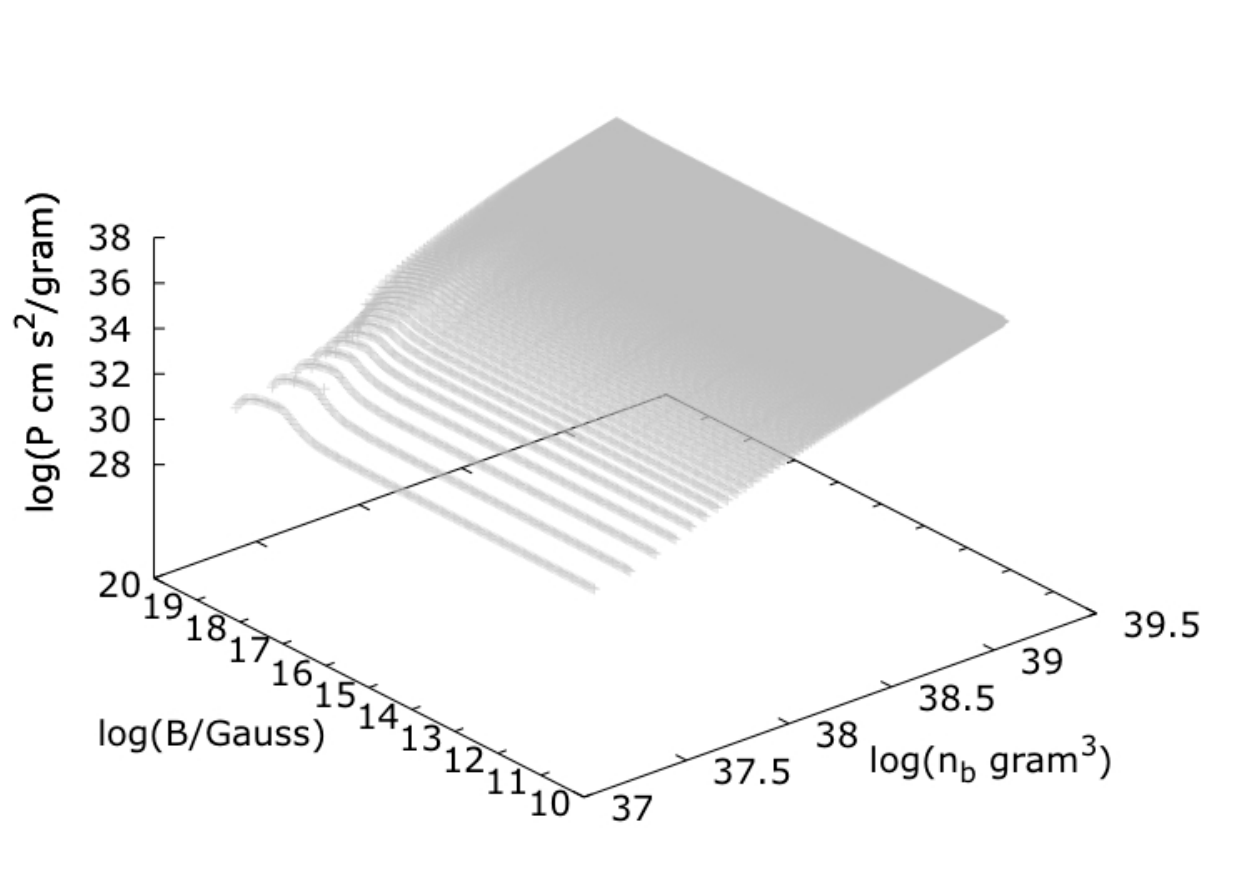}
\caption{Same as Fig. \ref{EOS30000cgs}, but now including the effect of anomalous magnetic moments. The magnetic field stiffens the EOS due to the polarization of the anomalous moments until complete spin polarization occurs ($B>10^{18}$ G). From that point on, the effect of Landau quantization takes over and softens the EOS.}
\label{EOS31000cgs}
\end{figure}

In this work, we employ the EOS calculated by \citet{magneticEOS}, which includes the effects of very strong magnetic fields in multicomponent, interacting matter. This is obtained via a field-theoretical approach in which the baryons (neutrons, $n$, and protons, $p$)  interact
via the exchange of $\sigma$-$\omega$-$\rho$ mesons. They consider two classes of models that differ in their high densities behavior. In these models, the effects of the anomalous magnetic moments of the nucleons are introduced via a covariant coupling of the baryons to the electromagnetic field tensor, presumed to be the result of of an appropriate high-energy theory. The  magnetic fields affect the EOS in two ways. In strong fields ($B>10^{14}$ G), the orbital motion of the charged particles is quantized (Landau quantization), resulting in a reduction in the electron chemical potential and a substantial increase in the proton fraction. This produces a softening of the EOS. However, if the anomalous moments are considered, the authors find that for very strong fields ($B>10^{18}$ G), complete spin polarization of the neutrons occurs. This produces an overall stiffening of the EOS that overwhelms the softening induced by Landau quantization. These effects are illustrated in Figures \ref{EOS30000cgs} and \ref{EOS31000cgs}.

\citet{magneticEOS2} have also studied the effects of hyperons on the magnetic EOS. Their EOS with hyperons, in the absence of a magnetic field is equivalent to the GM3 EOS calculated by \citet{GM3paper} where hyperons are introduced as free baryons, interacting only through weak interaction so that the system maintains nuclear statistical equilibrium. Typically, the introduction of the additional degrees of freedom associated with the new baryon species produces a net softening of the equation of state.

Since the pressure is now generally a function of both the matter density and magnetic field strength, explicit construction of a neutron star requires some knowledge about the magnetic field profile.

\section{A Tangled Magnetic Field}
\label{secB}
The primary processes responsible for the amplification of the magnetic field are believed to be dynamos driven by differential rotation and convection \citet{duncan92, thompson1993}.  These necessarily produce a tangled magnetic field inside the neutron star.  Since for simplicity we assume spherical symmetry throughout this work, we must model the dependence of the strength of this tangled magnetic field on radius. We do this approximately by making a handful of simplifying assumptions regarding the outcome of the dynamo process.

Various studies suggest that the dynamos naturally saturate at locally equipartition magnetic field strengths during the initial formation of the proto-magnetar after the supernova explosion, while different layers of the remaining stellar core condense to form the neutron star \citep{chevalier2005, naso2008, equipartition1}. That is, we expect
\beq
\epsilon_B \propto \epsilon_g,
\label{eps1}
\eeq
where $\epsilon_B$ is the local energy density of the magnetic field and  $\epsilon_g$ is the local energy density of the gas. Since $P_g\approx\epsilon_g/3$ at this time, this immediately implies
\beq
\beta \ {B^2 \over 8 \pi} = P_g,
\label{equiparteq}
\eeq
where $\beta$ is the standard proportionality factor relating the gas and magnetic pressures. 

As the proto-magnetar cools, $\beta$ will be approximately conserved.  During the cooling process the star will maintain hydrostatic equilibrium and hence for each spherical shell within the star
\beq
{P_g \over r} \sim \rho {G M \over r^2},
\eeq
where $r$ is the shell radius, $M$ is the enclosed mass, $\rho$ is the matter density, and $G$ is Newton's constant. Assuming that the magnetic field has grown sufficiently strong to suppress any further convection, $M$ will remain constant and the proto-magnetar will contract homologously. Thus $\rho$ will grow roughly proportional to $r^{-3}$ during the subsequent proto-magnetar evolution, and thus $P_g\propto r^{-4}$. Similarly, the flux conservation implies $B$ will grow $\propto r^{-2}$, and thus $B^2/8\pi\propto r^{-4}$ as well.  Therefore, we expect $\beta$ to remain fixed during the proto-magnetar's formation following the quenching of the dynamos.

\begin{figure}
\includegraphics[angle=0,scale=0.7]{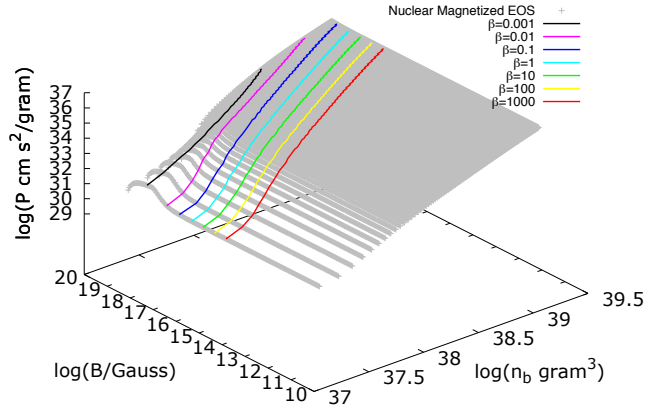}
\caption{EOS - pressure ($P$) versus number density of baryons ($n_b$) for different values of the magnetic field strength ($B$) assuming the existence of magnetic anomalous moments. The solid lines are the EOS found by imposing the equipartition condition (Equation \ref{equiparteq}) using different values of $\beta$. Smaller values of $\beta$ are associated with higher values of magnetic field strength.}
\label{plotequipart}
\end{figure}

This relation naturally provides a profile for the magnetic field which is proportional to the pressure of matter; given a global value of $\beta$ we may identify a unique pressure at each density. Examples of the EOS found by imposing Equation (\ref{equiparteq}) on the results of \citet{magneticEOS} assuming anomalous magnetic moments are shown in Figure \ref{plotequipart}.

In addition to the modification of the nuclear equation of state, the strong interior magnetic fields may produce magnetic stresses directly.  However, the nature of these stresses depend on the local geometry of the field as well as its strength.  Assuming that the outcome of the dynamo processes is a small-scale, tangled field that is weakly uncorrelated with the generating currents, the resulting stresses may be completely described by a magnetic pressure, which after assuming statistical isotropy is given by 
\beq
P_B= {1 \over 3} \  {B^2 \over 8\pi}. 
\label{pressure1}
\eeq
The factor of $1/3$ arises from the impact of magnetic tension; that it must be present is immediately evident from the fact that the electromagnetic stress-energy tensor is traceless. In Equation (\ref{equiparteq}), $\beta=1/3$ (giving the EOS lying between the cyan and blue curves in Figure \ref{plotequipart}), corresponds to a neutron star with equal pressure of magnetic field and matter. Smaller values of $\beta$, for which the pressure of the magnetic field will be larger than the one of matter, are not generically expected in turbulent MHD dynamos, as magnetic energy is converted back into thermal energy through magnetic reconnection. As we see in Figure \ref{plotequipart}, complete spin polarization of neutrons (due to their anomalous magnetic moments) occurs at $\beta =0.001$. Therefore, it is expected that at values of $\beta$ equal to, or larger than $1/3$, the anomalous magnetic moments will not have a considerable effect on the nuclear EOS and consequently in the M-R relations. We will see this explicitly in the next section. 

We note in passing that the assumptions made above regarding the correlation between the generating currents and the magnetic field need not be typically satisfied.  An extreme alternative are the force-free configurations described in \citet{narayan2008}, with a corresponding impact on the nature of the magnetic stresses.  However, we will postpone a detailed discussion of this until Section \ref{forcefree}. 

Before describing our M-R calculation, it is worth mentioning that the magnetized EOS of \citet{magneticEOS, magneticEOS2} describes the nuclear matter inside the neutron star. For the crust, i.e., densities below the nuclear saturation density (baryon rest-mass density of $\rho_{\rm ns} \sim 2.7 \times 10^{14}$ g cm$^3$), we use the SLy (Skyrme Lyon) EOS calculated by \citet{douchin2001}, which is based on the effective nuclear interaction SLy of the Skyrme type, which is useful in describing the properties of very neutron rich matter.






\section{M-R relations with an isotropic magnetic field}
\label{secMR}
As choosing a magnetic field configuration given by Equation (\ref{equiparteq}), allows us to calculate a unique magnetized EOS (as seen in Figure \ref{plotequipart}), we can now proceed to calculating M-R relations for neutron stars. As we are assuming spherical symmetry and isotropy, the Einstein equations  in hydrostatic equilibrium, simplify to the Tolman-Oppenheimer-Volkoff (TOV) equation:
\begin{multline}
{dP\over dr}= - {G\over r^2} \Big[ \epsilon+{P\over c^2}\Big] \Big[M+4\pi r^3 {P \over c^2}\Big]  \Big[1- {2GM \over c^2 r}\Big]^{-1},
\label{toveq}
\end{multline}
where $c$ is the speed of light, and $\epsilon$ and $P$ are the total energy density and pressure:
\bea
\epsilon=\epsilon_g + \epsilon_B, \\
P=P_g + P_B,
\eea
where $\epsilon_g$ and $P_g$ (the energy density and pressure of the gas), are related by imposing Equation (\ref{equiparteq}) (with a fixed choice of $\beta$), on the various magnetized nuclear EOS models of \citet{magneticEOS, magneticEOS2} (as shown in Figure \ref{plotequipart} for one particular nuclear EOS). Where stated we include the physics of the crust, by using the SLy EOS of \citet{douchin2001} below nuclear saturation density. The energy density of the magnetic field is given by $\epsilon_B=B^2/8\pi$, where again $B$ is fixed by the choice of $\beta$. The pressure of the magnetic field is given by Equation \ref{pressure1} (statistically isotropic force-inducing magnetic field). The mass enclosed inside radius $R$ is is related to local energy density by 
\beq
M(R)=\int_0^R \epsilon(r) \ 4\pi r^2 dr.
\label{MofR}
\eeq

\begin{figure}
\includegraphics[angle=0,scale=.40]{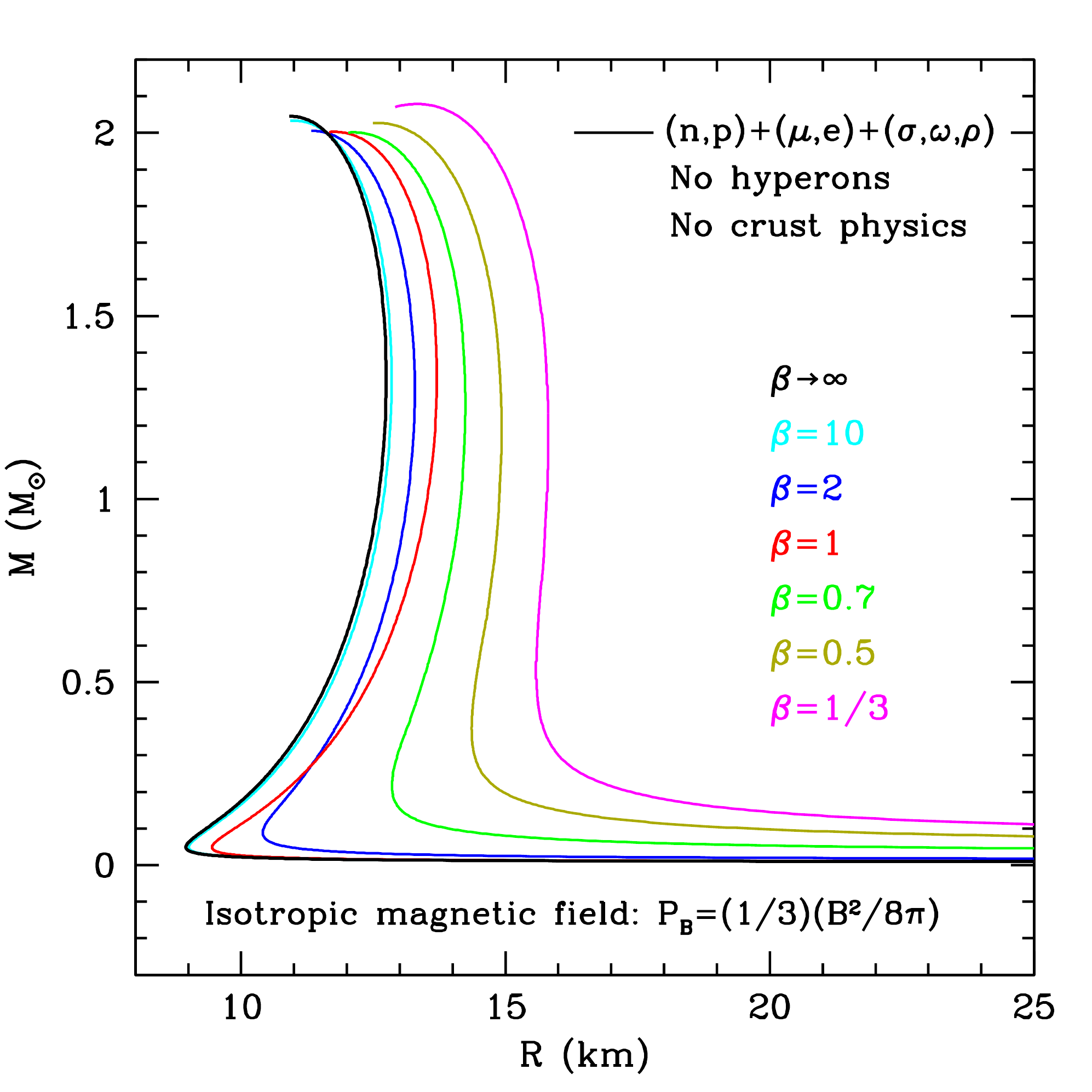}
\caption{The M-R relation for the magnetized EOS \citep{magneticEOS} with no crust physics included. Here, the magnetic field provides hydrostatic support for the star by entering globally in the Einstein equations as $P_B={(1/ 3)} (B^2 / 8\pi)$. The calculation has been done for different values of $\beta$ in Equation (\ref{equiparteq}). Smaller values of $\beta$ are associated with higher values of magnetic field strength in the neutron star. $\beta=1/3$ corresponds to a neutron star with equal pressure of magnetic field and matter.
\vspace{0mm}}
\label{MRglobalNOCRUST}
\end{figure}

We integrate Equation (\ref{toveq}) numerically, starting from the centre of the star and assuming different values of central pressure $P_g(r=0)$. Figure \ref{MRglobalNOCRUST} shows the M-R relations resulting from this integration in the absence of hyperons and crust physics.
Increasing the magnetic field strength (decreasing $\beta$) in the absence of hyperons has only a small effect on the neutron star maximum mass. However, larger magnetic field strengths are associated with larger neutron star radii.

\begin{figure}
\includegraphics[angle=0,scale=.40]{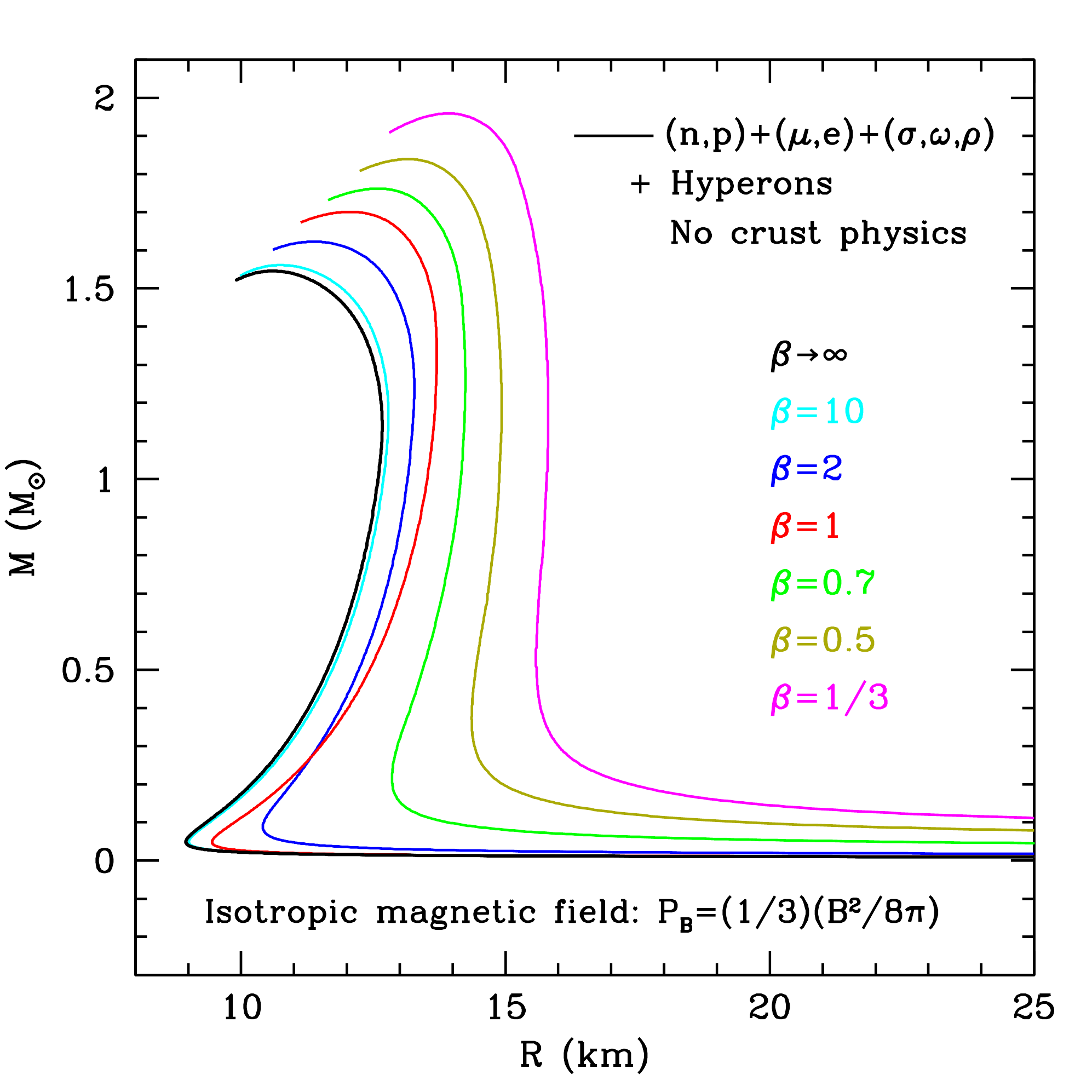}
\caption{Same as Figure \ref{MRglobalNOCRUST}, but now including the effect of hyperons. }
\label{MRglobalhyperonsNOCRUST}
\end{figure}

Figure \ref{MRglobalhyperonsNOCRUST} shows the M-R relations in the presence of hyperons, again ignoring the crust physics. In the presence of strong magnetic fields, all of the hyperons are susceptible to spin polarization. As we discussed in Section \ref{secEOS}, spin polarization counteracts the effects of Landau quantization, by stiffening the EOS. 
Although assuming the existence of hyperons gives a smaller maximum mass in the absence of a magnetic field ($\beta \rightarrow \infty$ in Figure \ref{MRglobalhyperonsNOCRUST}), increasing the magnetic field strength (decreasing $\beta$), now increases the maximum mass. Assuming equal pressures of magnetic field and matter ($\beta=1/3$), we get a maximum mass which is  26\% larger than the one with no magnetic field ($\beta \rightarrow \infty$).

\begin{figure}
\includegraphics[angle=0,scale=.40]{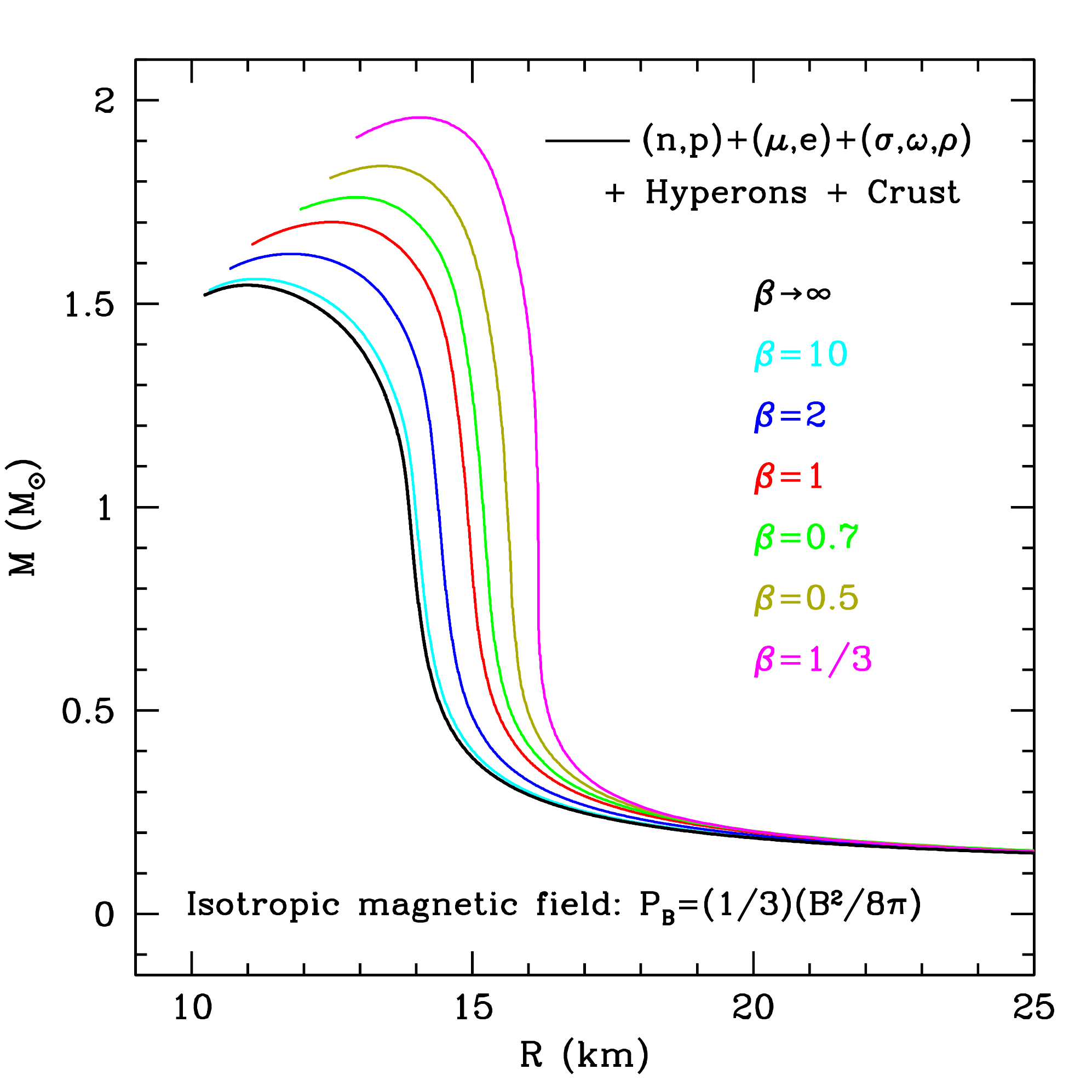}
\caption{Same as Figure \ref{MRglobalhyperonsNOCRUST}, but now also including the crust physics via the SLy EOS \citep{douchin2001}.  The black curve, corresponding to negligible magnetic field strength, is equivalent to the GM3 EOS \citep{GM3paper}.}
\label{MRglobalhyperons}
\end{figure}

Figure \ref{MRglobalhyperons} shows similar results, now incorporating the physics of the crust by using the SLy EOS calculated by \citet{douchin2001} at densities below nuclear saturation density. 
In this case, the M-R relation found by assuming zero magnetic field strength ($\beta \rightarrow \infty$), is similar to the well-known GM3 EOS calculated by \citet{GM3paper}. A similar increase in neutron star maximum mass with magnetic field strength is seen here.\footnote{For another recent study of the effects of strong magnetic fields on the population of hyperon stars, see \citet{gomes2014}.}

\begin{figure}
\includegraphics[angle=0,scale=.40]{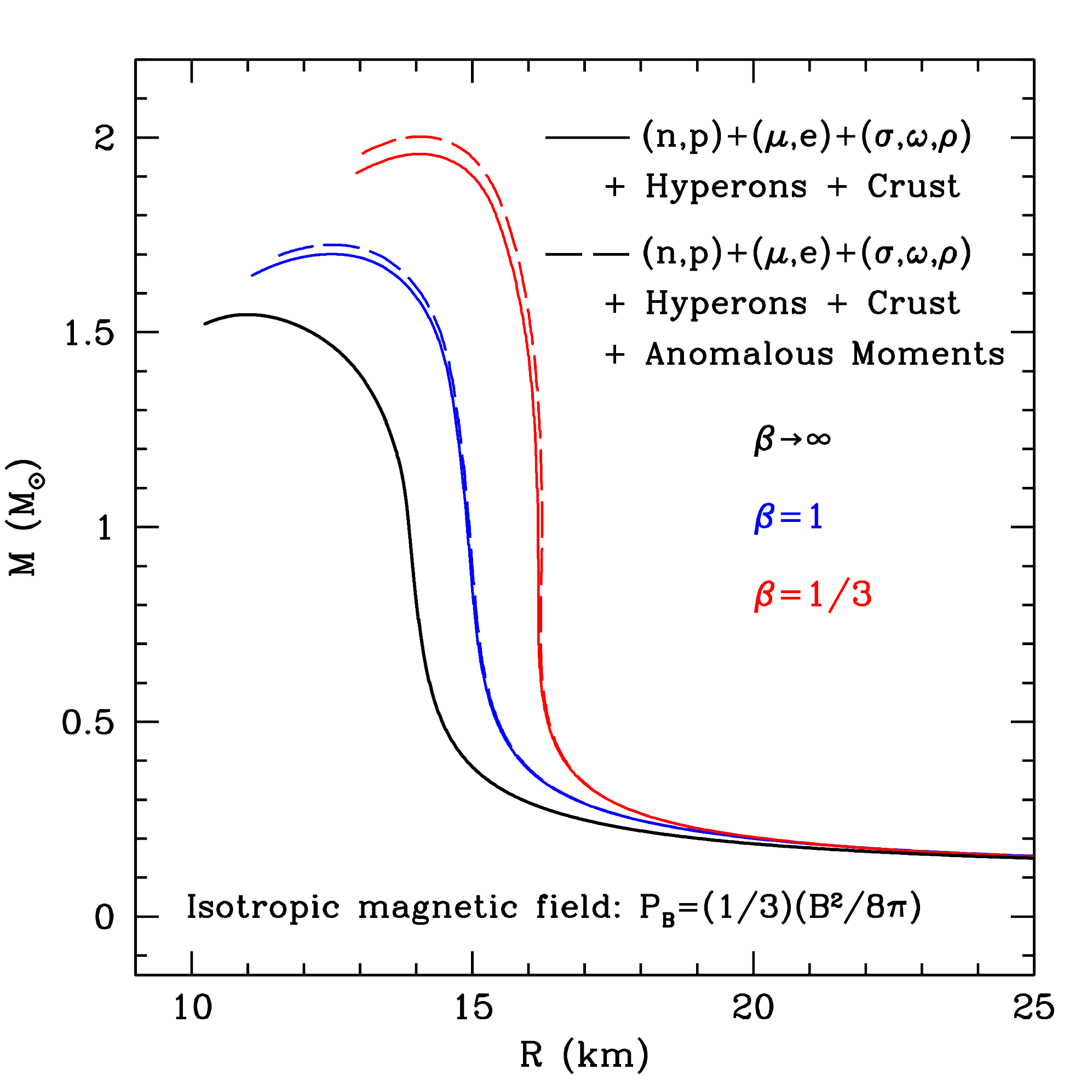}
\caption{The M-R relation for the magnetized EOS with hyperons  \citep{magneticEOS2} and the crust physics included via the SLy EOS \citep{douchin2001}. Here, the magnetic field provides hydrostatic support for the star by entering globally in the Einstein equations as $P_B={(1/ 3)} (B^2 / 8\pi)$. The solid curves have been calculated assuming no anomalous magnetic moments and the dashed curves have been calculated assuming the existence of anomalous moments. The calculation has been done for different values of $\beta$ in Equation (\ref{equiparteq}). Smaller values of $\beta$ are associated with higher values of magnetic field strength in the neutron star. $\beta=1/3$ corresponds to a neutron star with equal pressure of magnetic field and matter. The black curve, corresponding to negligible magnetic field strength, is equivalent to the GM3 EOS \citep{GM3paper}.}
\label{MRglobalhyperonsanomalous}
\end{figure}

We are also interested in the effects of the anomalous magnetic moments of nucleons on the M-R relations. Figure \ref{MRglobalhyperonsanomalous} shows these results for the magnetized nuclear EOS with hyperons and crust physics included. 
As we see, the effect of including the anomalous magnetic moments is small. We expected this from applying the condition given by Equation (\ref{equiparteq}) on the magnetic EOS of \citet{magneticEOS, magneticEOS2} with anomalous moments. As can be seen in Figure \ref{plotequipart}, complete spin polarization of the anomalous moments (which is where we would expect their effects to be considerable) occurs at values of $\beta \sim 0.001$, in other words when the pressure of the magnetic field is a 1000 times larger than pressure of matter, which are not expected to arise in dynamos, as discussed above. Therefore the strongest magnetic field strength we considered was one for which the magnetic pressure equals the pressure of matter ($\beta=1/3$). The effect of including magnetic anomalous moments under this magnetic field strength is a 2\% change in the neutron star maximum mass as seen in Figure \ref{MRglobalhyperonsanomalous}. 

\begin{figure}
\includegraphics[angle=0,scale=.40]{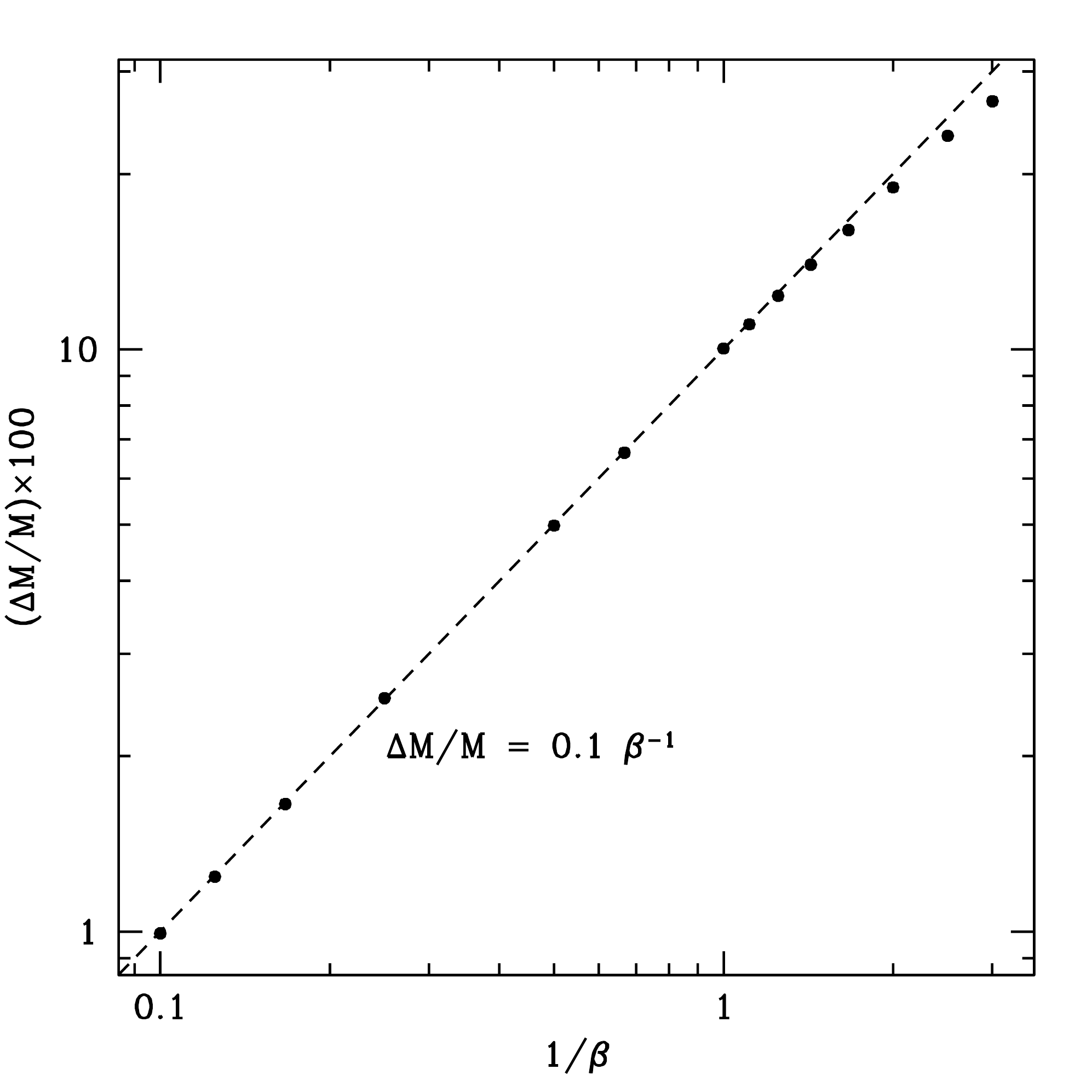}
\caption{Change in maximum mass of neutron stars as a function of $1/\beta$ (see Equation (\ref{equiparteq}) for definition of $\beta$) for the mass-radius relation shown in Figure \ref{MRglobalhyperons}.}
\label{MaxMbeta}
\end{figure}

\begin{figure}
\includegraphics[angle=0,scale=.40]{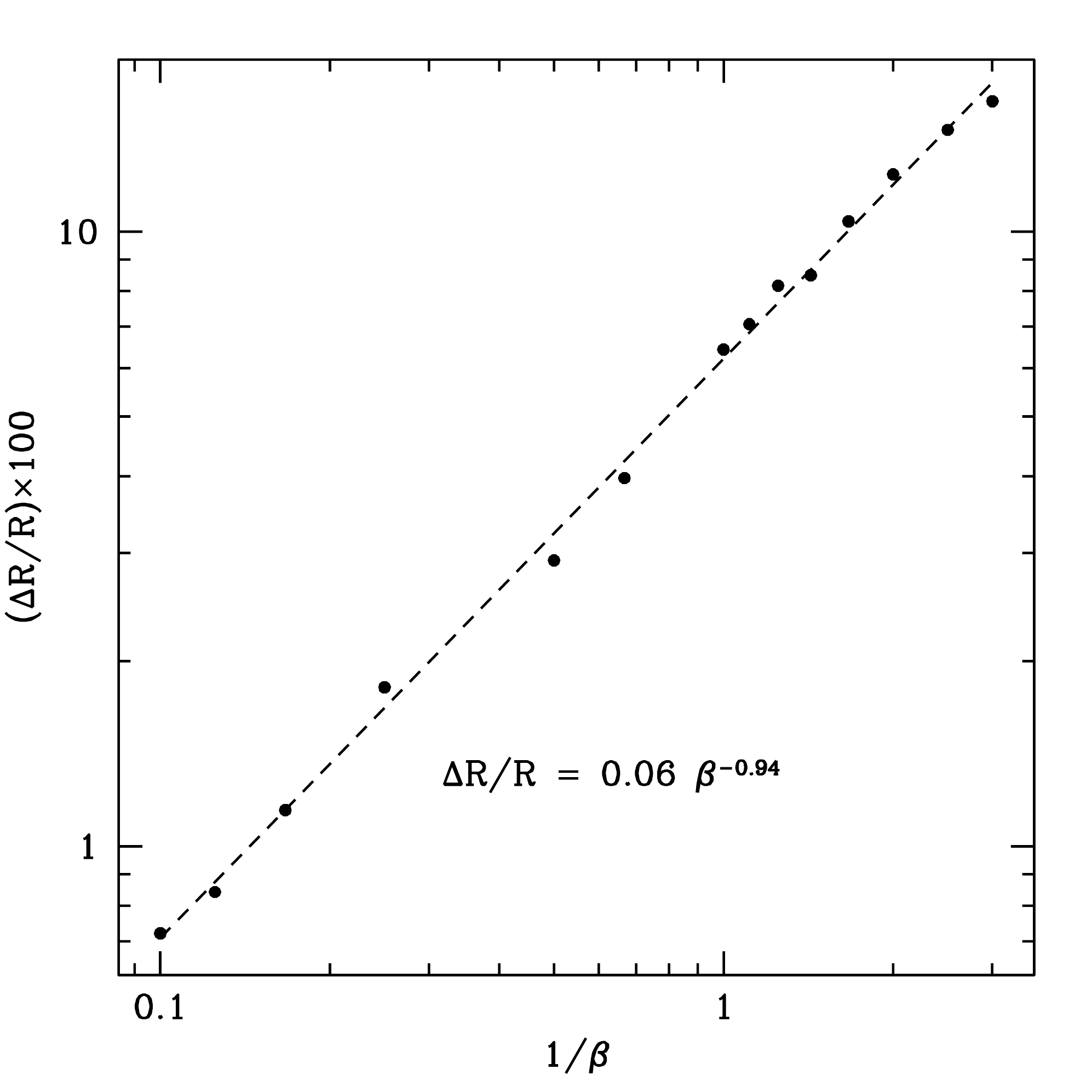}
\caption{Change in radius of neutron stars as a function of $1/\beta$ (see Equation (\ref{equiparteq}) for definition of $\beta$) for the mass-radius relation shown in Figure \ref{MRglobalhyperons}.}
\label{InfRbeta}
\end{figure}

Finally, to better illustrate the effect of the magnetic field strength on the maximum mass and radius of neutron stars, Figures \ref{MaxMbeta} and \ref{InfRbeta} show the maximum mass and inflection point in radius (where the second derivative of mass versus radius changes sign) as a function of $1/\beta$, for the mass radius relationship shown in Figure \ref{MRglobalhyperons}. We specify the radius by the inflection point in the M-R relation, where mass is nearly independent of radius, giving a characteristic value across a wide range of neutron star masses. We see that these relative changes are well fit by
\beq
\frac{\Delta M_{\rm max}}{M_{\rm max}} \simeq 0.1 ~\beta^{-1}, ~{\rm and} ~\frac{\Delta R_{\rm inf}}{R_{\rm inf}} \simeq 0.06~ \beta^{-0.94},
\eeq
for an {\it isotropic} tangled magnetic field distribution. 


\section{M-R relations with a force-free magnetic field}
\label{forcefree}

Despite the production of small-scale, turbulent fields by the dynamos initially, the final magnetic field configuration remains unclear.  After the dynamos quench and before the formation of a crust that can balance shear stresses the neutron star will reconfigure the magnetic field geometry via bulk fluid motions on a timescale measured in Alfv\'en crossing times, resulting in a linked, nearly force-free geometry dictated by the initial magnetic helicity \citep{2004Natur.431..819B,2006A&A...450.1097B}.  This is a natural consequence of helicity conservation during reconnection, corresponding to the minimum energy state at fixed magnetic helicity \citep{narayan2008}, which is fully defined by
\beq
\nabla\times\textbf{B} = \alpha\textbf{B}\,,
\eeq
from which it is clear that $\textbf{j}\times\textbf{B}=0$ where $\textbf{j}$ is the current density, i.e., the Lorentz force vanishes.  In the above, $\alpha$ is a constant set by the boundary conditions at the stellar surface, setting a scale length for the resulting magnetic geometries.

This gives axisymmetric and anisotropic configurations for the magnetic field. This configuration in the interior of the star, in terms of the vector spherical harmonics $\bY_l$,  $\bPsi_l$ and  $\bPhi_l$ is:
\begin{multline}
\textbf{B} = \sum_l \left\{
-\left[\frac{l(l+1)}{r} d_l j_l(\alpha r)\right] \bY_l\right.\\
-\left[\frac{1}{r}\partial_r r d_l j_l(\alpha r)\right] \bPsi_l
\left.+ \alpha d_l j_l(\alpha r) \bPhi_l
\right\},
\label{eq:ffB}
\end{multline}
where $l$ is the spherical harmonic degree, $j_l$ is the spherical Bessel function, and $d_l$ is a constant giving the strength of each harmonic mode.

To assess the consequence of the force-free condition on the hydrostatic equilibrium of the star, we can start by looking at the Newtonian Euler equation and impose static conditions:
\beq
\rho {d \textbf{v} \over dt} = - \nabla P_g - \rho \nabla \Phi +  \textbf{j} \times  \textbf{B} = - \nabla P_g - \rho \nabla \Phi  =0,
\eeq
where $\rho$ is the density and $\Phi$ is the Newtonian potential given by the Possion equation:
\beq
\nabla^2 \Phi = 4 \pi G \rho.
\eeq
Therefore, we have:
\beq
\nabla P_g = -\rho \nabla{\Phi},
\eeq
which is the Newtonian equation for hydrostatic equilibrium with no effect of the magnetic field in it. This means that in Newtonian mechanics, the force-free condition implies that the magnetic field will play no global role in supporting the star, even though it might still change the nuclear equation of state via the effects mentioned above (e.g., Landau quantization, spin polarization, etc.).

In general relativity, the force-free condition can be written in its covariant form:
\beq
j_{\nu} F^{\mu \nu} =0 \rightarrow F^{\mu \nu} \nabla_\alpha F_\nu^\alpha=0,
\eeq
where $F_{\mu \nu}$ is the electromagnetic tensor. The electromagnetic stress-energy tensor is given by
\beq
4 \pi {T_{EM}}^\mu_\nu = F^{\mu \alpha}F_{\nu \alpha} - {1\over 4} \delta^\mu_\nu F^{\alpha \beta} F_{\alpha \beta}.
\eeq
The full stress-energy tensor is given by:
\beq
T^{\mu \nu} ={T_{gas}}^{\mu \nu} + {T_{EM}}^{\mu \nu},
\eeq
from which the Euler equation is obtained by $\nabla_\nu T^{\mu \nu} =0$. The contribution from ${T_{EM}}^{\mu \nu}$ is generally, after some manipulation and application of Maxwell's equations,
\beq
\nabla_\nu{T_{EM}}^{\mu \nu} ={1\over 4 \pi} F^{\mu \alpha} \nabla_\nu F^\nu_\alpha = j_\nu F^{\mu \nu} =0,
\label{decoupled}
\eeq
therefore as in the Newtonian limit the direct impact of the electromagnetic forces vanish, i.e., $\nabla_\nu{T}^{\mu \nu} = \nabla_\nu{T_{gas}}^{\mu \nu} =0$.

Gravity is now described by the Einstein equation:
\beq
G^{\mu \nu}=8\pi G \big({T_{gas}}^{\mu \nu}+{T_{EM}}^{\mu \nu}\big).
\eeq
Unlike the Newtonian case, now the electromagnetic field can contribute by sourcing gravity. In practice, it is necessary to solve for the full magnetic field configuration and metric simultaneously. However, we might imagine that since gravity is a long-range force and the structure of the initially highly turbulent electromagnetic field should exhibit mostly small-scale structure that we may spatially average the electromagnetic configuration. That is to solve:
\beq
G^{\mu \nu}=8\pi G \big({T_{gas}}^{\mu \nu}+\langle{T_{EM}}^{\mu \nu}\rangle \big),
\label{approxeinstein}
\eeq
where $\langle ... \rangle$ is a local spatial average, i.e., an average over scales large in comparison to the scale of the electromagnetic field fluctuations and small in comparison to the gravitational scales. The stress energy tensor for this average field is 
\[\langle{T_{EM}}^{\mu \nu}\rangle = \left( \begin{array}{cccc}
-\epsilon_B&  &  &  \\
 & P^{(r)}_B &  & \\
 &  &  P^{(\theta)}_B & \\
 &  &  & P^{(\phi)}_B
\end{array} \right),\]
where 
\begin{eqnarray*}
\epsilon_B&=& \langle B^2 \rangle/8\pi,\\
P^{(r)}_B&=&  \langle B^2 \rangle/8\pi - \langle B^r B_r \rangle/4\pi,\\
P^{(\theta)}_B&=&  \langle B^2 \rangle/8\pi -  \langle B^\theta B_\theta\rangle /4\pi,\\
P^{(\phi)}_B&=&  \langle B^2 \rangle/8\pi -  \langle B^\phi B_\phi\rangle /4\pi.
\end{eqnarray*}
The anisotropy of this average configuration may be quantified via an anisotropy parameter, 
\beq
\Delta \equiv {[P^{(\theta)}_B + P^{(\phi)}_B]/2 - P^{(r)}_B \over  \langle B^2 \rangle /8\pi}.
\label{deltadefine}
\eeq
An anisotropy parameter $\Delta=-1$ corresponds to a purely tangential magnetic field configuration $\big(\langle B^r \rangle=0 \big)$ and $\Delta=2$ corresponds to a purely radial field \big($\langle B^r \rangle=\langle B \rangle \big)$. An isotropic average field will have $\Delta=0$.

\begin{figure}
\includegraphics[angle=0,scale=.45]{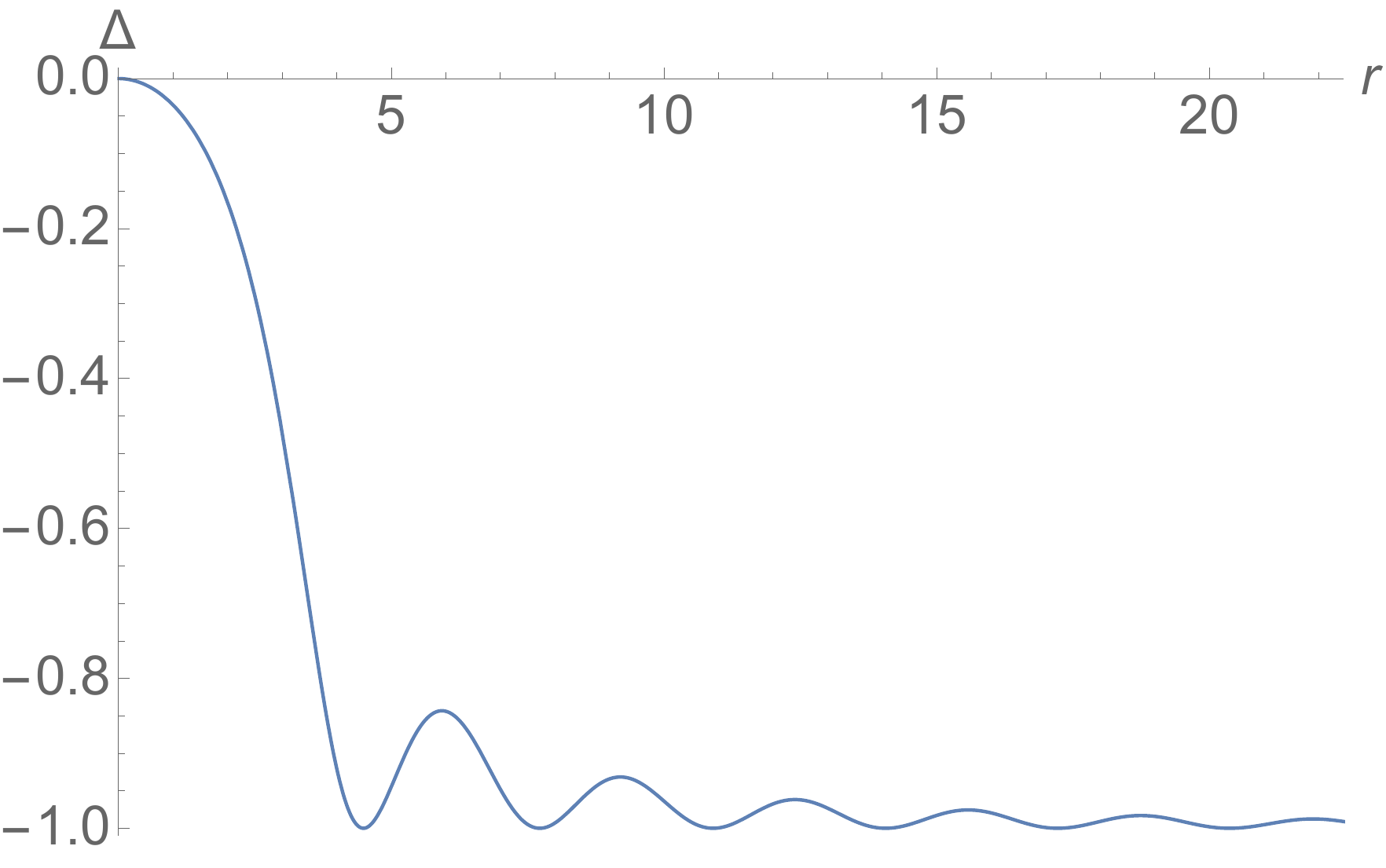}
\caption{Anisotropy parameter for the average force-free configuration for the $l=1$ spherical harmonic degree. Units of radius are set so that $\alpha=1$ in Equation (\ref{eq:ffB}).}
\label{Anisol=1}
\end{figure}
\begin{figure}
\includegraphics[angle=0,scale=.45]{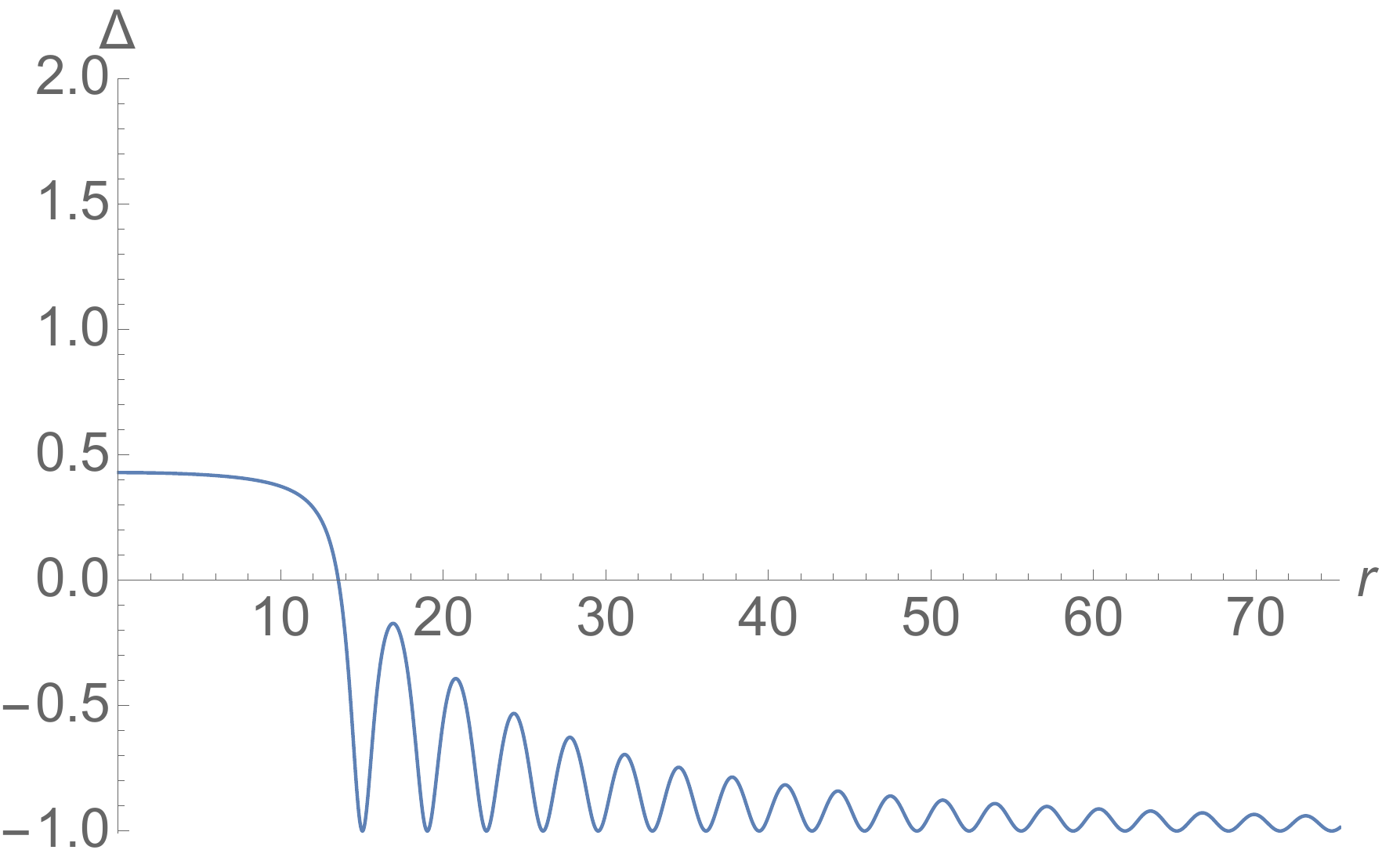}
\caption{Anisotropy parameter for the average force-free configuration for the $l=10$ spherical harmonic degree. Units of radius are set so that $\alpha=1$ in Equation (\ref{eq:ffB}).}
\label{Anisol=10}
\end{figure}

Figures \ref{Anisol=1} and \ref{Anisol=10} show the anisotropy parameter calculated for the average force-free magnetic field configuration given by Equation (\ref{eq:ffB}) for two spherical harmonic degrees of $l=1$ and $l=10$. At the centre of the star, there is a region with an isotropic average field ($\Delta = 0$) for $l=1$ and as we go to higher spherical harmonic degrees $\Delta \rightarrow 0.5$ at the centre as can be seen in Figure \ref{Anisol=10}. As we see in both Figures \ref{Anisol=1} and \ref{Anisol=10}, $\Delta$ drops to -1 towards the outside of the star. This matches our expectation of field lines closing and the magnetic field becoming tangential in the outside. In the calculations below, we will refer to the radius of the isotropic core with $\Delta = 0$ as $R_\star$ (having chosen the anisotropy parameter of the first harmonic degree, as the magnetic field strength is dominated by the first harmonic degree at the centre).

To calculate the mass-radius relationship of the star, we will assume spherical symmetry and continue using an equipartition magnetic field (Equation (\ref{equiparteq})).  In order to write the hydrostatic equilibrium equations coming from general relativity for the force-free model, we use Equation (\ref{decoupled}) (in other words the fact that the Euler equation is simply given by the energy momentum conservation of the gas). As an approximation, we will solve the Einstein equation given by Equation (\ref{approxeinstein}). It is worth clarifying that solving Equation (\ref{approxeinstein}) alongside the energy momentum conservation of the gas ($\nabla_\nu{T_{gas}}^{\mu \nu} =0$), would formally require
\beq
\nabla_\mu \langle {T_{EM}}^{\mu \nu} \rangle =0,
\label{conserv1}
\eeq
while by using the force-free condition Equation (\ref{decoupled}) we are only assuming that 
\beq
\nabla_\mu {T_{EM}} ^{\mu \nu}=0.
\label{conser2}
\eeq 
As we will show in the Appendix, solving Equation (\ref{conserv1}) alongside with the Euler equation for the gas, would uniquely determine the anisotropy parameter. As spherical symmetry only allows isotropy ($\Delta=0$) at the centre of the star, solving Equation (\ref{conserv1}) will result in $\Delta \rightarrow 0$ throughout most of the star. This will change the final mass-radius relationship results we obtain from using the anisotropy parameter coming directly from the force-free configuration (satisfying Equation (\ref{conser2})) at most by 9\%. This anisotropy parameter based on the results shown in Figure \ref{Anisol=1} will have a core with $\Delta=0$ of size $R_\star$ and in our approximation, $\Delta$ will drop to $-1$ at $R_\star$.

In dimensions where $G=1$ and $c=1$, the Euler equation coming from $\nabla_\nu{T_{gas}}^{\mu \nu} =0$ is given by 
\beq
{dP_g \over dr} = -(\epsilon_g +P_g) \  g,
\label{eulergas}
\eeq
where $g$ is 
\beq
g\equiv {M+4\pi \left(P_g+P_B^{(r)}\right) r^3 \over r^2 (1-2M/r)}.
\label{gTOV}
\eeq
The enclosed mass is still given by Equation (\ref{MofR}), which still includes the magnetic contribution to the energy density. 

\begin{figure}
\includegraphics[angle=0,scale=.4]{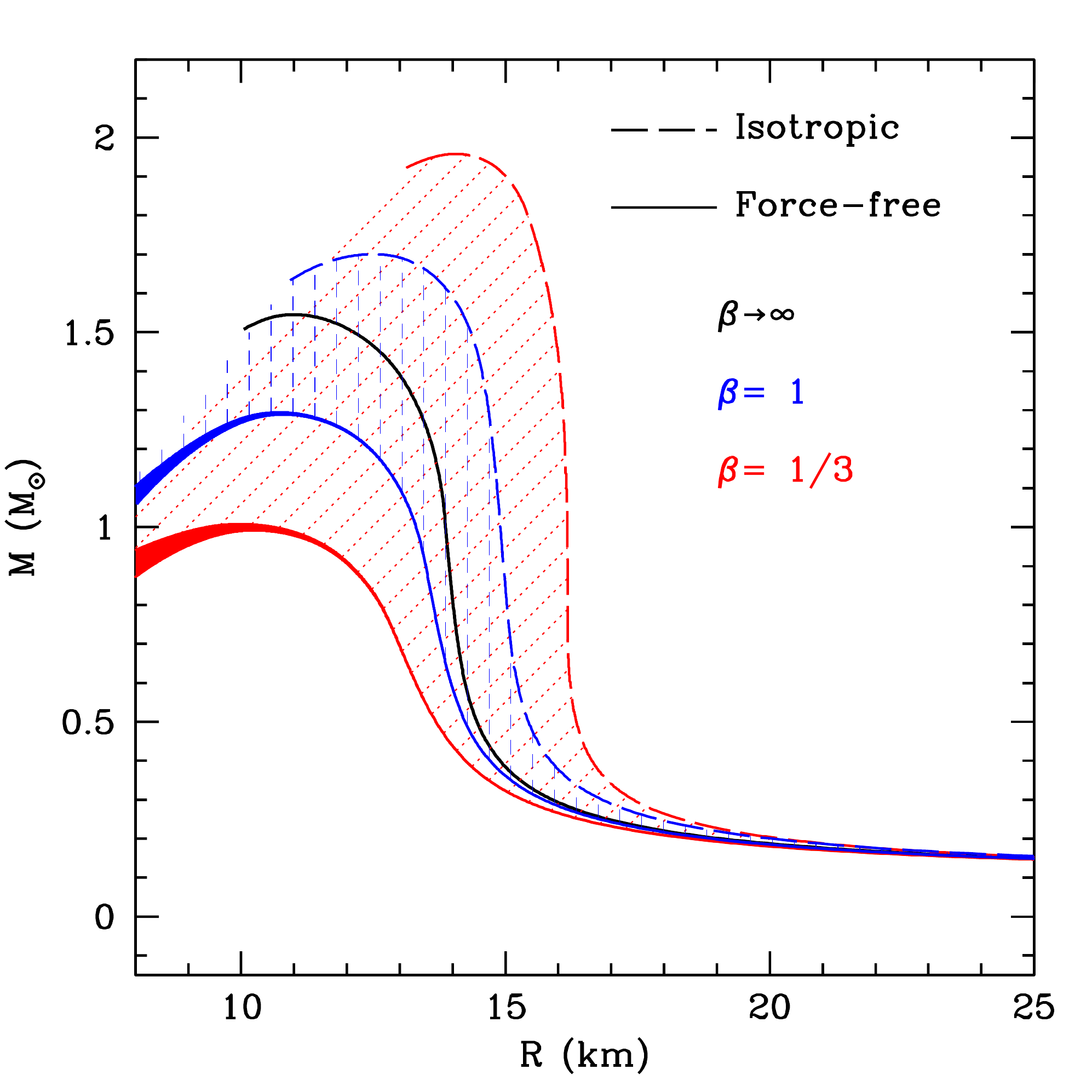}
\caption{The M-R relations for the magnetized EOS with hyperons  \citep{magneticEOS2} and the crust physics included via the SLy EOS \citep{douchin2001}. The dashed curves are for an isotropic magnetic field and the the solid curves correspond to the force-free model. Results are shown for two values of $\beta$. The range in the force-free (solid) curves is due to assuming a range in $R_\star$ (radius of the isotropic core with $\Delta=0$) from $0$ to $3$ km. }
\label{Force-free+IsotropicMR}
\end{figure}

Based on the definition of $\Delta$ and the assumption of spherical symmetry $\langle B^\theta B_\theta\rangle=\langle B^\phi B_\phi\rangle $, from which 
\beq
P_{B}^{(r)} =   {1-2\Delta \over 3} \ \langle B^2 \rangle/8\pi.
\label{pbr}
\eeq
Using this and the equipartition assumption the total radial pressure becomes 
\beq
P_g+P_B^{(r)} = (1+ {1-2 \Delta \over 3 \beta}) \ P_g,
\label{ptot}
\eeq
requiring only the specification of $P_g$ and $\Delta$.
The pressure and energy density of the gas are related through the nuclear equation of state which is itself affected by the magnetic field. The resulting mass-radius relationships from this numerical integration are shown in Figure \ref{Force-free+IsotropicMR}.
The solid curves correspond to the force-free mass-radius calculation based on the magnetized EOS with hyperons  \citep{magneticEOS2} and the crust physics included via the SLy EOS \citep{douchin2001}. This calculation has been done for two different values of $\beta$. The range in each curve is due to assuming a range in $R_\star$ (radius of the isotropic core with $\Delta=0$) from $0$ to $3$ km. The previously found mass-radius relationships based on an isotropic magnetic field are also plotted for comparison (dashed curves). 

In stark contrast to the case of tangled, isotropic magnetic fields, a force-free configuration produces a substantial {\em decrease} in the maximum neutron star mass.  This is not surprising -- force-free magnetic field geometries contribute to the energy density, and therefore gravity, but not to the supporting pressure.  This conclusion is only weakly dependent on the details of the force-free configuration under consideration, showing little sensitivity to the size of our assumed isotropic core, $R_\star$.  However, even for large values of $\beta$, e.g., $\beta=1$, the magnitude of the effect is constrained to 20\%, similar in size if not direction to that associated with the isotropic tangled geometry.

\begin{figure}
\includegraphics[angle=0,scale=.40]{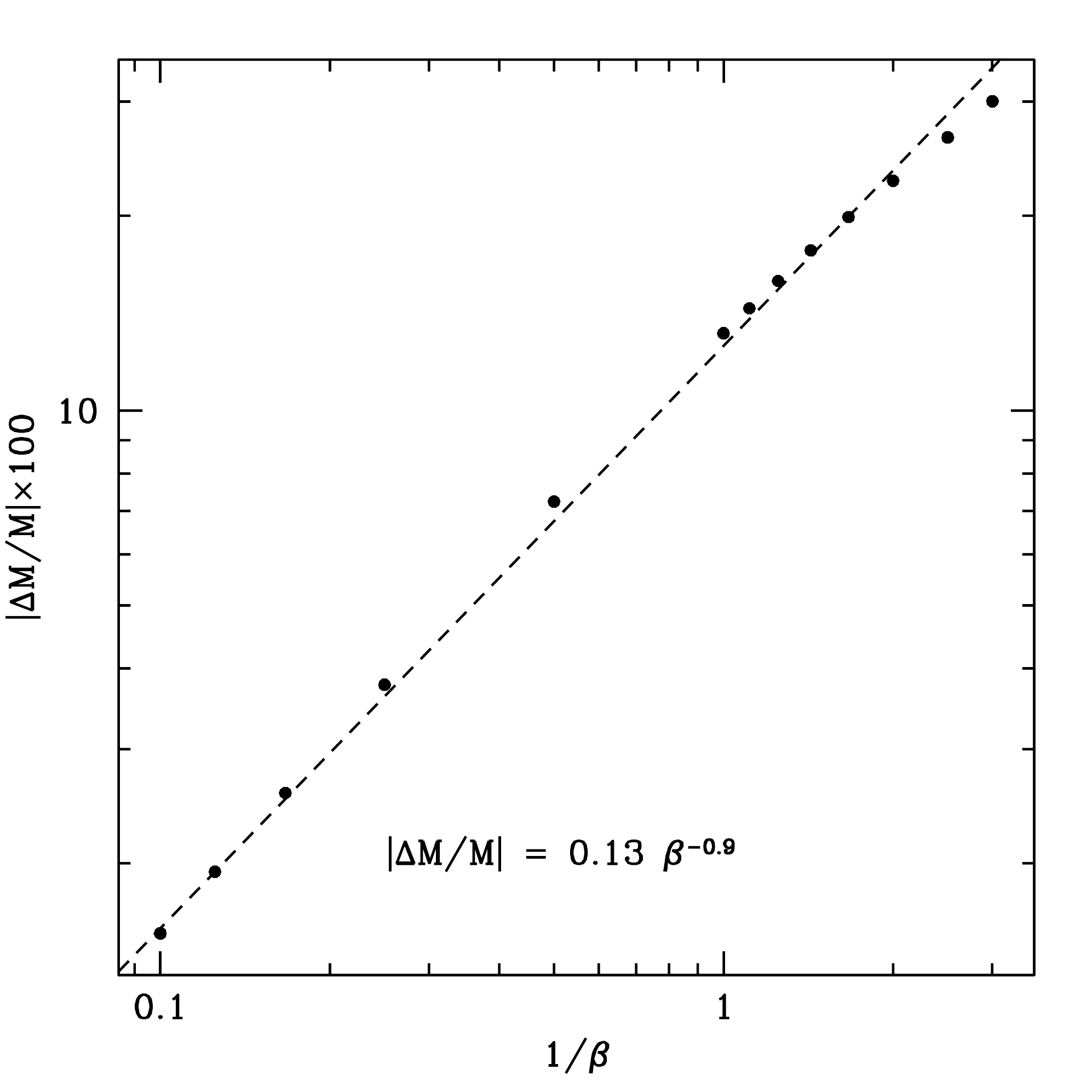}
\caption{Change in maximum mass (absolute value for the logarithmic scale as the mass is decreasing) of neutron stars as a function of $1/\beta$ (see Equation (\ref{equiparteq}) for definition of $\beta$) for the force-free mass-radius relations shown in Figure \ref{Force-free+IsotropicMR}.}
\label{MaxMbetaforcefree}
\end{figure}

\begin{figure}
\includegraphics[angle=0,scale=.40]{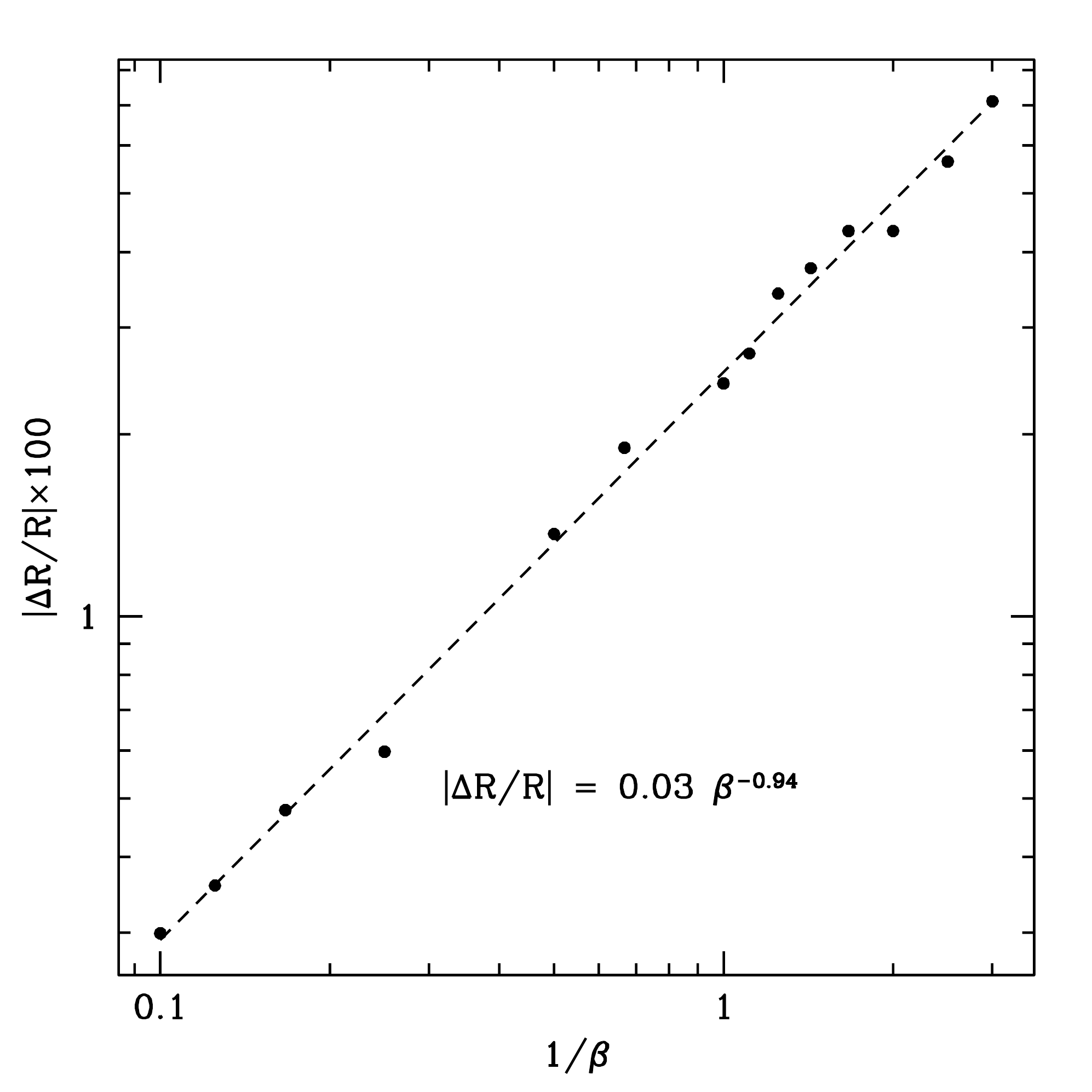}
\caption{Change in radius of neutron stars (absolute value for the logarithmic scale as the radius is decreasing) as a function of $1/\beta$ (see Equation (\ref{equiparteq}) for definition of $\beta$) for the force-free mass-radius relations shown in Figure \ref{Force-free+IsotropicMR}.}
\label{InfRbetaforcefree}
\end{figure}

The dependence on magnetic field strength of the size of the effects on mass and radius are illustrated in Figures \ref{MaxMbetaforcefree} and \ref{InfRbetaforcefree}.
Note that unlike Figures \ref{MaxMbeta} and \ref{InfRbeta}, the change in mass and radii are now negative.  As before, the change in mass and radii are well fit by power laws over at least an order of magnitude in $\beta$, for maximum mass and inflection point radius
\beq
\frac{\Delta M_{\rm max}}{M_{\rm max}} \simeq -0.13 ~\beta^{-0.9}, ~{\rm and} ~\frac{\Delta R_{\rm inf}}{R_{\rm inf}} \simeq -0.03~ \beta^{-0.94},
\eeq
assuming a {\it force-free} tangled magnetic field distribution.













\section{Conclusions}
\label{conclude}
In this paper, we have demonstrated that the assumed geometry of interior neutron star magnetic fields is at least as important as the field strengths themselves when assessing the impact on masses and radii, capable of {\em qualitatively} modifying their effect.  Highly tangled, isotropic magnetic configurations do result in an increase in the maximum mass, typical of results often found in the literature \citep[e.g., see][]{astashenok2014}.  However, force-free configurations, motivated by dynamical studies of magnetic field evolution in proto-neutron stars, produce a {\em decrease} in the maximum neutron star mass.  For both configurations of the equipartition strength fields, the magnitude of the impact on the maximum mass is limited to 30\%.  The impact of anomalous magnetic moments for equipartition magnetic field strengths at best modify the maximum mass by 2\%, suggesting that they are sub-dominant to the global magnetic field geometry.

This immediately challenges the assertion that magnetic fields provide a means to reconcile the recent observations of very massive neutron stars with unmagnetized equations of state whose maximum mass is otherwise precluded.  In the absence of some credible mechanism for producing super-equipartition magnetic fields, the maximum deviation in the maximum mass of order 20\%-30\% means that at best these can impact only marginal cases.  For example, this is unable to bring the GM3 EOS we employed into consistency with the existing $2~M_\odot$ mass measurements.  We expect this to be generic for models of neutron star formation in which the interior magnetic fields are produced by turbulent processes, providing an optimistic upper limit on the mass-limit enhancement achievable via the introduction of electromagnetic support.  Of course, the existence of well-motivated magnetic field configurations in which the maximum neutron star mass actually decreases (i.e., force-free configurations) calls into question the value of invoking magnetic support altogether in order to increase maximum neutron star masses.

We note that the observed anomalous braking indices for the spin-down of pulsars can be interpreted as evidence for sub-surface toroidal magnetic fields of $10^{14}-10^{15}$ G, {\it substantially} stronger than the observed dipole fields, which are slowly diffusing out of the crust due to transport processes \citep[e.g., see][and references therein]{2015MNRAS.446.1121G}. Even if taken at face value, it is unclear if these fields can reach the viral level of $\sim 10^{18}$ G in the core. However, as we have argued above, this is anticipated by simple astrophysical dynamo processes during the formation of the neutron star, followed by subsequent cooling in (quasi-)hydrostatic equilibrium. 

An interesting possibility is that, depending on their initial conditions, different neutron stars might have dynamos of varying efficiency during their formation process. Therefore, future observations of Mass-Radius relationship of neutron stars may exhibit an intrinsic scatter due to variations in internal magnetic field. An even more exciting possibility is finding a surface measure of this internal magnetic field (e.g., through modelling the braking indices), which could then correlate with (and thus effectively reduce) this scatter, following our simple scalings (Figs.~\ref{MaxMbeta}-\ref{InfRbeta},\ref{MaxMbetaforcefree}-\ref{InfRbetaforcefree}). In particular, the sign of the correlation will be indicative of the (force-free v.s. isotropic) field configuration.     

In typical MHD turbulence, and in lieu of an active dynamo, one expects the magnetic field to relax to a force-free configuration within an Alfv\'en crossing time, through subsequent reconnections that conserve helicity \citep{2006A&A...450.1097B,narayan2008}.  Therefore, one may expect the force-free configuration to be preferred. However, if the interior of neutron stars transitions into type II superconductor, as hypothesized in the literature \citep[e.g., see][and references therein]{2013ApJ...764L..25H}, then it can support a frustrated network of flux tubes, which do {\it not} reconnect \citep{1995RPPh...58.1465B}. Such a configuration can support tangled (statistically) isotropic magnetic fields. 

Finally, one may further speculate that reconnection is not completely halted, but is only significantly slowed (e.g., due to the presence of a mixture of type I and II superconducting phases in the interior). Therefore, the field configuration will slowly transition from isotropic to force-free, leading to a reduction in the maximum mass. 
Via this mechanism a subset of isolated neutron stars, those with masses between those that can be supported by force-free and tangled magnetic field configurations, could undergo a delayed collapse, surviving for a time scale set by the evolution of the internal magnetic geometry.  Such an event may present an attractive alternate candidate for energetic, short-time scale phenomena, e.g., short gamma-ray bursts and the recently detected fast radio bursts.

\acknowledgments

We thank Ramesh Narayan and Chiamaka Okoli for very fruitful discussions. This work was supported by the Natural Science and Engineering Research Council of Canada, the University of Waterloo and by Perimeter Institute for Theoretical Physics. Research
at Perimeter Institute is supported by the Government of Canada through Industry
Canada and by the Province of Ontario through the Ministry of Research \& Innovation.

\bibliography{References}

\appendix
\label{appen}

As mentioned in Section \ref{forcefree}, solving the average Einstein equation (\ref{approxeinstein}):
\beq
G^{\mu \nu}=8\pi G \big({T_{gas}}^{\mu \nu}+\langle{T_{EM}}^{\mu \nu}\rangle \big),
\eeq
along with the energy momentum conservation of the gas:
\beq
\nabla_\nu{T_{gas}}^{\mu \nu} =0,
\eeq
requires:
\beq
\nabla_\mu \langle {T_{EM}}^{\mu \nu} \rangle =0.
\label{EMconserv}
\eeq
This equation results in the anisotropic TOV  equation for the average magnetic field:
\beq
{dP_B^{(r)} \over dr} = -(\epsilon_B + P_B^{(r)})  \ g + {2 \  \Delta \  \langle B^2 \rangle/8\pi \over r},
\label{TOVmagnetic}
\eeq
where $g$ is given by Equation (\ref{gTOV}). As $P_B^{(r)}$ is given by Equation (\ref{pbr}), and assuming Equation (\ref{equiparteq}), we will have:
\begin{eqnarray}
{d \over dr} \Big[ {P_g \over \beta} ({1-2\Delta \over 3}) \Big] \ &=& \ -\Big[ {P_g \over \beta} + {P_g \over \beta} ({1-2\Delta \over 3})  \Big] \ g + {2 \Delta P_g  \over \beta r} \\
{1-2\Delta \over 3 \beta} \  {dP_g \over dr} - {2 \over 3}{P_g \over \beta}  {d \Delta \over dr} \ &=& \  2 P_r  {\Delta -2 \over 3 \beta} \ g  +  {2 \Delta P_g  \over \beta r},
\end{eqnarray}

and substituting for $dP_g / dr$ from Equation (\ref{eulergas}), and solving for $d \Delta / dr$, we get:
\beq
{d \Delta \over dr} = \Big[ {3 \over 2} + (\Delta - {1 \over 2}) {\epsilon_g \over P_g} \Big] \ g - {3 \Delta \over r},
\eeq

which simplifies to:
\beq
r^3 \Delta = \int_0^r  \Big[ {3 \over 2} + (\Delta - {1 \over 2}) {\epsilon_g \over P_g} \Big] \ g \  {r}^3 dr.
\label{deltaTOV}
\eeq

\begin{figure}
\centerline{\includegraphics[angle=0,scale=0.8]{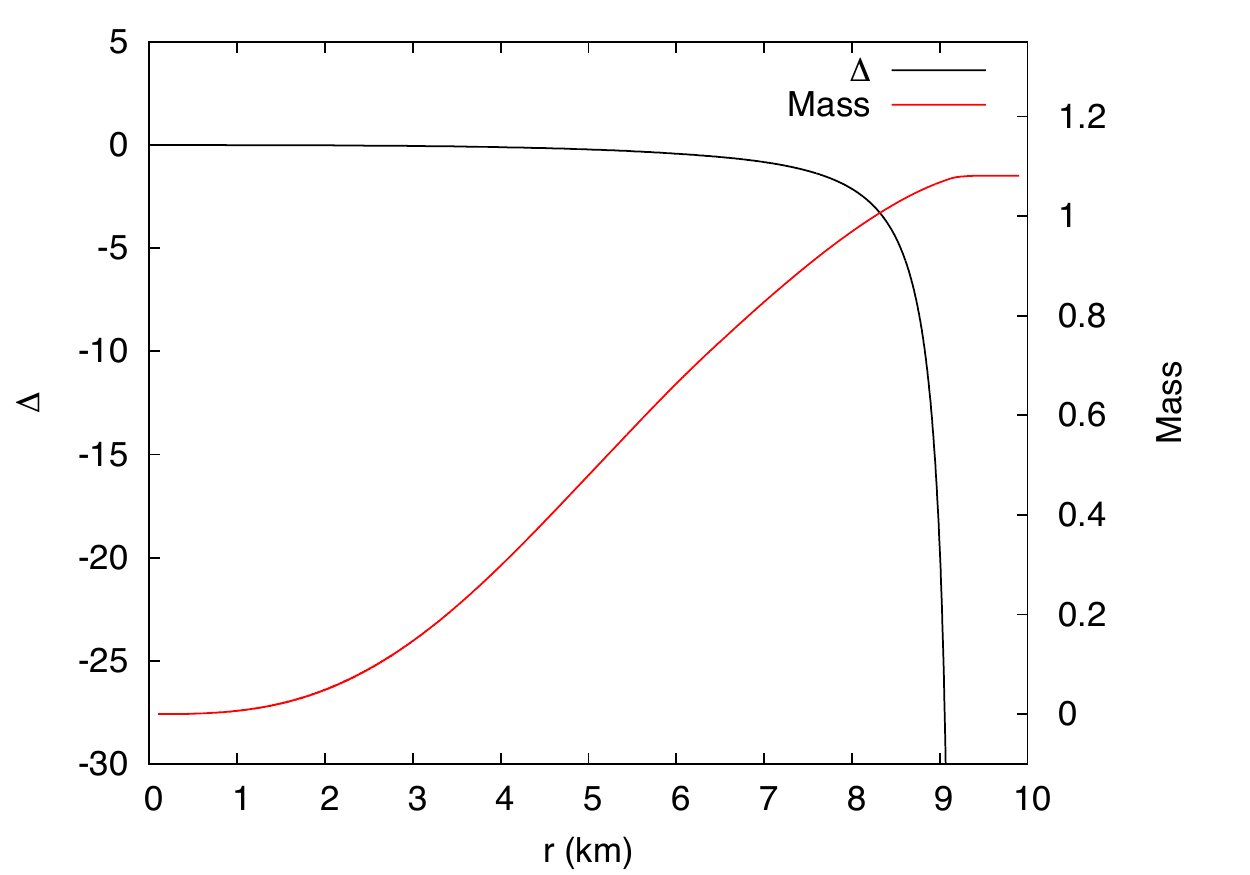}}
\caption{$\Delta(r)$ for a neutron star of radius $R=10$ km with $\beta =1/3$. $\Delta$ remains very close to zero (isotropic field) and more than $-1$ until the edge of the star where it drops to lower values. }
\label{delta}
\end{figure}

\begin{figure}
\centerline{\includegraphics[angle=0,scale=0.4]{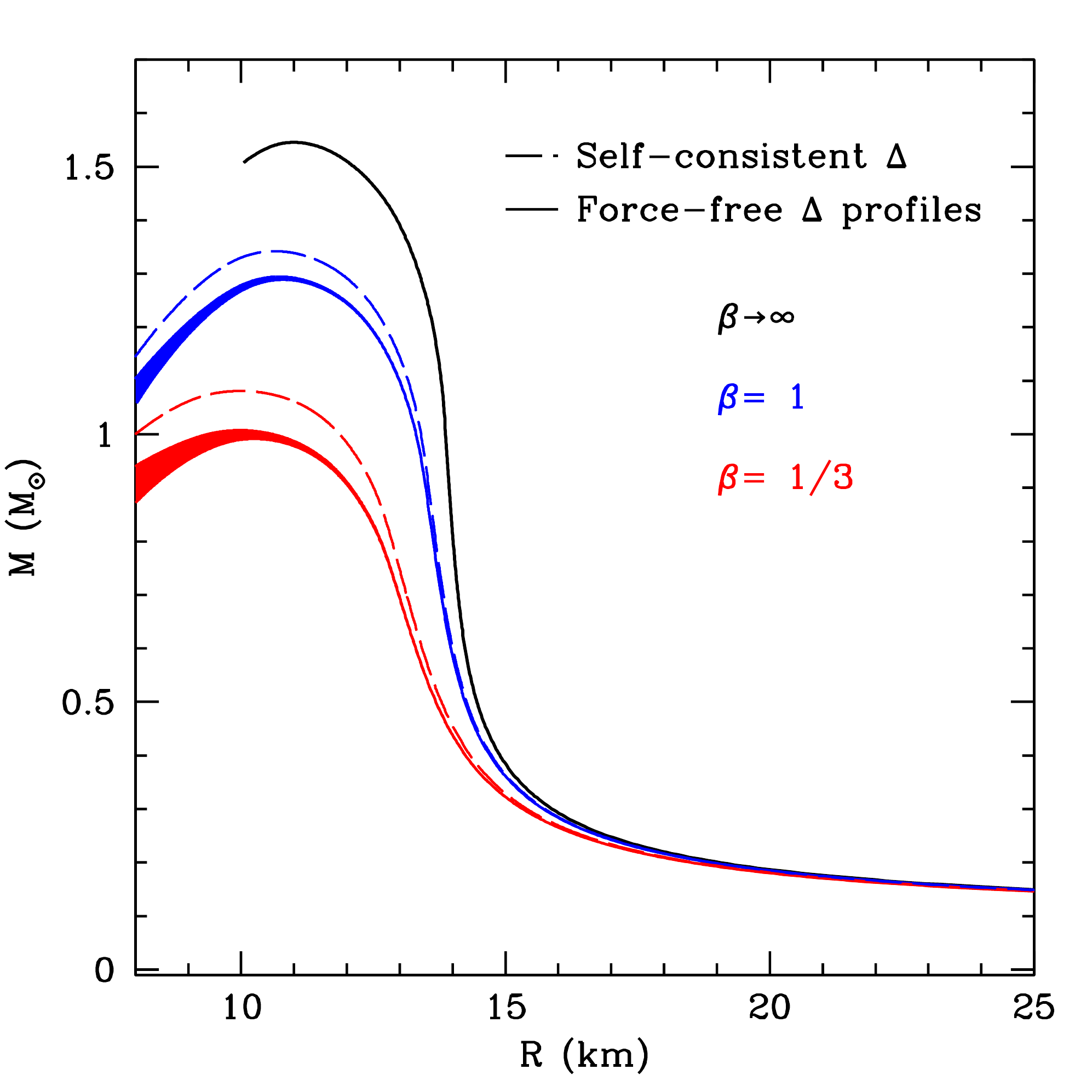}}
\caption{Solid curves are the force-free mass-radius relationships already shown in Figure \ref{Force-free+IsotropicMR}. The dashed curves are the results found for the self-consistent anisotropy calculation presented in the Appendix.}
\label{anisocompare}
\end{figure}

The suitable boundary condition for $\Delta$ is determined by the fact that in the anisotropic TOV equation for the magnetic field (\ref{TOVmagnetic}), the second term on the right side is well defined only if $\Delta \rightarrow 0$ when $r \rightarrow 0$. Equation (\ref{deltaTOV}) solved along with the TOV equation for the gas, uniquely determines $\Delta(r)$. As we see in Figure \ref{delta}, $\Delta$ remains very close to zero (isotropic field) and more than $-1$ until the edge of the star where it drops to lower values. This behaviour of $\Delta$ makes at most a 9\% change (in case of $\beta=1/3$ and smaller changes for larger values of $\beta$) to the mass radius relationships previously found as can be seen in Figure \ref{anisocompare}. It is important to note that values of $\Delta<-1$ are unphysical, implying imaginary magnetic field strengths.  This raises some concern regarding the self-consistent solution near the stellar surface; for the case shown in Figure \ref{delta}, roughly 20\% of the mass is located in this unphysical regime, implying that the mass estimates from the self-consistent solutions are marginal over-estimates.

\end{document}